\documentclass[aps,twocolumn,pra,showpacs]{revtex4-1}
\usepackage{graphicx,color}
\usepackage{amsmath}
\usepackage{amssymb}
\usepackage{epstopdf}
\begin{document}
\title{Entanglement detection on an NMR quantum information
processor using random local measurements }
\author{Amandeep Singh}
\email{amandeepsingh@iisermohali.ac.in}
\affiliation{Department of Physical Sciences, Indian
Institute of Science Education \& 
Research (IISER) Mohali, Sector 81 SAS Nagar, 
Manauli PO 140306 Punjab India.}
\author{Arvind}
\email{arvind@iisermohali.ac.in}
\affiliation{Department of Physical Sciences, Indian
Institute of Science Education \& 
Research (IISER) Mohali, Sector 81 SAS Nagar, 
Manauli PO 140306 Punjab India.}
\author{Kavita Dorai}
\email{kavita@iisermohali.ac.in}
\affiliation{Department of Physical Sciences, Indian
Institute of Science Education \& 
Research (IISER) Mohali, Sector 81 SAS Nagar, 
Manauli PO 140306 Punjab India.}
\begin{abstract}
Random local measurements have recently been
proposed to construct entanglement witnesses and
thereby detect the presence of bipartite
entanglement.  We experimentally demonstrate the
efficacy of one such scheme on a two-qubit NMR
quantum information processor.  We show that a set
of three random local measurements suffices to
detect the entanglement of a general two-qubit
state.  We experimentally generate states with
different amounts of entanglement, and show that
the scheme is able to clearly witness
entanglement.  We perform complete quantum state
tomography for each state and compute state
fidelity to validate our results.  Further, we
extend previous results and perform a simulation
using random local measurements to optimally
detect bipartite entanglement in a hybrid system
of $2\otimes 3$ dimensionality.
\end{abstract}
\pacs{03.67.Bg,03.67.Lx,03.67.Mn}
\maketitle
\section{Introduction}
\label{intro}
Entanglement is an important resource for quantum
information processing, and many  questions about
its characterization,  optimal detection  and its
protection from decoherence are still
open~\cite{horodecki-rmp-09}.  The classification
and detection of entanglement is a demanding task
and attempts to do so, have relied on methods
including those based on Bell-type
inequalities~\cite{huber-pra-11,jungnitsch-prl-11,wallman-pra-12},
quantum state
tomography~\cite{white-prl-98,thew-pra-02},
dynamic learning tools and numerical
schemes~\cite{behrman-qic-13,spedalieri-pra-12},
entanglement
witnesses~\cite{lewenstein-pra-00,guhne-jmo-03,arrazola-pra-12},
positive partial transpose
mixtures~\cite{novo-pra-13,bartkiewicz-pra-15-1},
and expectation values of Pauli
operators~\cite{zhao-pra-13,miranowicz-pra-14}.
The negativity under partial transpose (NPT) is a
necessary and sufficient condition for existence
of entanglement in $2\otimes2$ and $2\otimes3$
dimensional quantum
systems~\cite{horodecki-pla-96,peres-prl-98}. For
higher dimensional systems, although a number of
sufficient conditions in terms of entanglement
witnesses are available, the problem of complete
characterization of entanglement is still
open~\cite{guhne-pr-09}.  Experimental
explorations of entanglement in the context of NMR
quantum information processing, include
implementation of an entanglement
witness~\cite{filgueiras-qip-12} and measurement
of bipartite quantum correlations of an unknown
quantum state~\cite{silva-prl-13}.  Pseudo-bound
entanglement was experimentally generated and
ground-state entanglement studied in a system of
three NMR
qubits~\cite{suter-pra-10,suter-pra-10-1}.
Three-qubit entanglement was characterized on an
NMR quantum information
processor~\cite{shruti-pra-15,debmalya-pra-15} and
the evolution of multiqubit entangled states was
studied with a view to control their
decoherence~\cite{kawamura-ijqc-06,harpreet-pra-14}.
It is important, particularly from an experimental
point of view, to be able to detect entanglement
by a minimum number of experiments, requiring as
little effort as possible.

A promising direction of research in the detection
of quantum entanglement has been the use of local
observables to find an optimal decomposition of
entanglement witnesses~\cite{guhne-pra-02}.  The
method assumes some prior knowledge of the density
matrix and is able to unequivocally detect
entanglement by performing only a few local
measurements~\cite{guhne-ijtp-03,toth-prl-05,guhne-njp-10}.
These entanglement detection schemes have been
recently extended to the case of completely
unknown states with no prior information~\cite{dagmar-njp-15}.  
This scheme uses a set of random local measurements
and optimizes over the space of possible 
entanglement witnesses that can be constructed
thereof.

This work focuses on experimentally using a set of random
local measurements to detect bipartite entanglement of
unknown pure entangled states.   Our experiments demonstrate
the optimality of using random local measurements to detect
entanglement in a system of two qubits on an NMR quantum
information processor.  We obtain the expectation values of
a set of local measurement operators and use semi-definite
programming to thereby construct the witness operator to
detect the presence of entanglement.  We show that a set of
three local measurements are sufficient to unequivocally
detect entanglement of most entangled states of two qubits.
We experimentally generate several states with different
amounts of entanglement and evaluate their entangled (or
separable) nature by performing this optimal set of local
measurements.  We validate these results by constructing
experimental tomographs of each state and compute negativity
as a measure of entanglement from them.  With a view to
generalize these methods to larger spaces, we perform a
simulation to detect bipartite entanglement of unknown pure
entangled states in a $2 \otimes 3$ system, using a set of
random measurements acting locally on the qubit and the
qutrit involved. It is observed that by performing a
few measurements, the entanglement of most states gets
implicated. 

The material in the paper is arranged as follows:
in Section~\ref{expt} we first present the
semi-definite programming based protocol to detect
entanglement, and later describe its experimental
implementation on two NMR qubits.  In
Section~\ref{hybrid} we present the results of
simulations on a $2 \otimes 3$ system.
Section~\ref{concl} contains a few concluding
remarks.
\section{Experimental detection of entanglement 
using local measurements}
\label{expt}
In typical experiments to detect entanglement
based on witnesses,  knowledge about the state
is required  beforehand. One may argue that if the
state is already known or has been tomographed,
then one can calculate its entanglement
properties by using the witness.  In
the present study we take a different approach, where
we do not ask for \textit{a priori} state
information, but instead strategically choose the
local measurements.  Semi-definite programming
(SDP) is then used to obtain the relative weights
of the expectation values of these local
measurements which are then used to  build the
entanglement witness for the unknown state.

We followed the procedure outlined by Szangolies
et.~al.~\cite{dagmar-njp-15} to construct a class
of decomposable entanglement witness operators for
an unknown state using random local measurements.
In this protocol, once we fix the set of
measurements, the witness is optimized to increase
the possibility of detecting entanglement. 
Consider a composite system 
in a Hilbert space
${\cal{H}}_{AB}={\cal{H}}_{A}\otimes
{\cal{H}}_{B}$. 
A witness operator is a Hermitian operator $W$
in this composite space, such that ${\rm Tr}(W\rho) \geq 0\quad$ 
for all
separable $\rho$ and ${\rm Tr}(W\rho) \neq 0$ for
at least one entangled $\rho$. A witness is called
decomposable if it can be written in terms of two
positive operators $P$ and $Q$ such that
\begin{equation} 
W= P + Q^{{\rm T}_A}.
\end{equation} 
where the operation ${\rm T}_A$
represents the partial transpose with respect
to subsystem $A$.  Further,
since we would like to build the witness operator
out of local measurements, consider local
Hermitian operators 
$A_i \in {\cal{H}}_A$ and 
$B_j \in {\cal{H}}_B$ where $i,j$ are decided by
the number of measurements that we are wish to
carry out for each local system. We would
therefore like the witness operator to be given as 
\begin{equation}
W=\sum_{i,j}c_{ij}A_i\otimes B_j
\label{block_decomposition}
\end{equation}
with $c_{ij} \in \cal{R}$. It should be noted here
that if we allow $A_i$ and $B_j$ to run over a
complete set of bases in the local operator spaces,
then by Bloch decomposition, every Hermitian
operator can be written in the form given in
Eqn.(\ref{block_decomposition}) as
described in~\cite{dagmar-njp-15}. However, in
our case we will first choose the measurements
that we want to experimentally perform and then optimize
the witness operator in such a way that we
maximize the chances of detecting entanglement.
Finding the expectation value of 
the entanglement witness operator $W$ 
(given the set of local observables $A_i$ and $B_j$)
is equivalent to finding the coefficients $c_{ij}$
subject to the trace constraints on
the witness operator.
Let us define a column vector $\bf{c}$ where we
take the columns of $c_{ij}$ and stack them one
below the other, and similarly define a vector
$\bf{m}$ in which we stack the experimentally measured
expectation values $\langle A_i \otimes
B_j\rangle$ as a long column vector such that
\begin{equation}
{\rm Tr} (W \rho)  = 
\sum_{i,j} c_{ij} \langle
A_i\otimes B_j \rangle=
\bf{c}.\bf{m}
\end{equation}
The SDP looks for the class of
entanglement witness operators with unit trace and
decomposable as $ P+Q^{T_A} $. This decomposition ensures
the detection of bipartite NPT states.
The corresponding SDP can be
constructed as~\cite{dagmar-njp-15}:
\begin{eqnarray*}
{\rm Minimize :~} &&\bf{c}.\bf{m}\\
\mbox{Subject to}~:
&& {\rm ~(1)~~} W=\sum_{i,j}c_{ij}(A_i\otimes B_j)\\
&&{\rm ~(2)~~} W=P+Q^{T_A}\\
&&{\rm ~(3)~~} P\geq 0\quad {\rm and} \quad
Q\geq 0\\
&&{\rm ~(4)~~} Tr(W)=1
\end{eqnarray*}
We implemented the SDP using
MATLAB~\cite{matlab-book} subroutines that
employed SeDuMi~\cite{sturm-oms-99} as an SDP
solver.

We now turn to the problem of experimentally
measuring the expectation values of various observables
using NMR, 
for a system of $N$ weakly
interacting spin-1/2 particles. The density
operator for this system can be decomposed as a linear combination of
products of Cartesian spin angular momentum 
operators $I_{ni}$, $n$ labeling the spin and
$i=x,y$ or $z$~\cite{ernst-book}.
For two qubits 
a total of 16
product operators
completely span the space of all $4 \times 4$ Hermitian
matrices.  
The four maximally
entangled Bell states for two qubits, 
and their corresponding 
entanglement witness operators, can always be written as a
linear combination of the three product operators
$2 I_{1x}I_{2x}$, $2 I_{1y}I_{2y}$, $2 I_{1z}I_{2z}$ and the 
identity operator.  
We use the symbols 
$O_i ( 1 \le i \le 15)$  to represent these
product operators, with
the first three symbols $O_1$, $O_2$ and
$O_3$ representing the operators $2 I_{1x}I_{2x}$,
$2 I_{1y}I_{2y}$ and $2 I_{1z}I_{2z}$ respectively.  
We need to experimentally determine the expectation values 
of these operators $O_i$,
for the state $\rho$ whose entanglement is to
be characterized.  
The expectation
values of these 
operators are mapped to the local $z$-magnetization of 
either of the two qubits by specially crafted
NMR pulse sequences, and are summarized in Table~I.
The expectation values 
are obtained by measuring  the
$z$-magnetizations of the corresponding qubit.
The NMR pulse sequences given in Table~I transform
the state via a single measurement, which is completely
equivalent to the originally intended measurement of
local operators, and considerably simplifies the
experimental protocol.
\begin{figure}[h]
\includegraphics[scale=1]{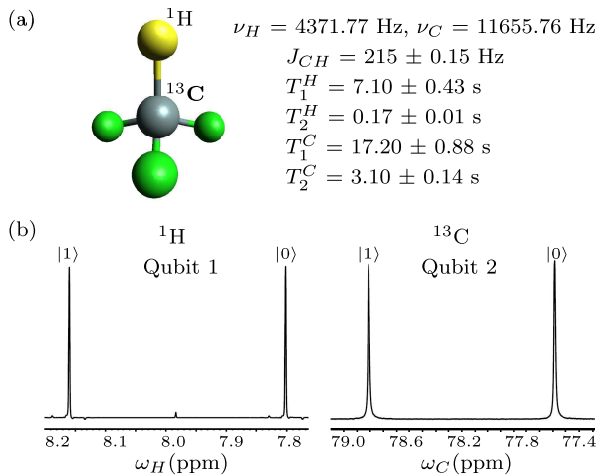}
\caption{(Color online) (a) 
Structure of the ${}^{13}$C enriched chloroform 
molecule with the two qubits labeled as ${}^{1}$H and
${}^{13}$C.  Tabulated experimental NMR parameters 
with chemical shifts ($\nu_i$) and spin-spin
coupling ($J_{ij}$) in Hz, and relaxation times
$T_1$ and $T_2$ in seconds,
and (b) ${}^{1}$H and ${}^{13}$C NMR spectra obtained
at thermal equilibrium after a $\frac{\pi}{2}$ readout
pulse. The spectral resonances of each qubit are labeled by the logical
state $\{ \vert 0 \rangle, \vert 1 \rangle \}$ of the
other qubit.}
\label{mol}
\end{figure}

The two NMR qubits were encoded in a molecule of ${}^{13}$C
enriched chloroform, with the ${}^{1}$H and ${}^{13}$C
nuclei being labeled as the first and the second
qubit, respectively. The molecular structure, experimental 
parameters, 
and the NMR spectrum  of the thermal
initial state
are shown in Fig.\ref{mol}. Experiments were
performed at room temperature on a Bruker Avance III 600 MHz
NMR spectrometer equipped with a QXI probe. 
The Hamiltonian of this weakly interacting two-qubit system
in the rotating frame~\cite{ernst-book} is
\begin{equation}
{\cal H} = \nu_{H} I^{H}_{z} + \nu_{C} I^{C}_{z}
+ J_{CH} I^{H}_{z} I^{C}_{z}
\end{equation}
where $\nu_{H}, \nu_{C}$ are the Larmor resonance
frequencies and $I^{H}_{z}, I^{C}_{z}$ are the $z$ components
of the spin angular momentum operators for the proton and
carbon nuclei respectively, and $J_{CH}$ is the spin-spin
coupling constant.

The two-qubit system was initially prepared in the
pseudopure state $\vert 00 \rangle$ using the 
spatial averaging technique~\cite{cory-physicad}, with the
density operator given by
\begin{equation}
\rho_{00} = \frac{1}{4}(1- \epsilon) I + \epsilon \vert 00 \rangle 
\langle 00 \vert
\end{equation}
where $\epsilon \approx 10^{-5}$ is an estimate of the
thermal polarization.
We note here that
NMR is an ensemble technique that can 
experimentally observe only
deviation density matrices (with zero trace).
The state fidelity was calculated from the 
Uhlmann-Jozsa relation~\cite{uhlmann-rpmp-76,jozsa-jmo-94} 
\begin{equation}
F=\left(Tr\left(\sqrt{\sqrt{\rho_{th}}\rho_{ex}
\sqrt{\rho_{th}}}\right)\right)^2
\end{equation} 
where $\rho_{th}$ and $\rho_{ex}$ represent
the theoretical and experimentally prepared density matrices
respectively. 
The experimentally prepared pseudopure state was
tomographed using full quantum state 
tomography~\cite{leskowitz-pra-04} and the state
fidelity was computed to be $ 0.99 \pm 0.01$.

\begin{figure}[h]
\includegraphics[scale=1]{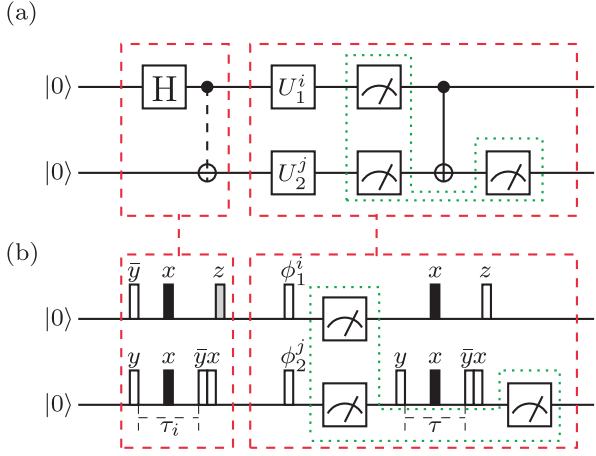}
\caption{(Color online)
(a) Quantum circuit to implement the entanglement detection
protocol.  The first red box creates states with different
amounts of entanglement.
The second red box maps the observables
$O_i$ to the $z$-magnetization of either qubit. Only one
$z$-magnetization is finally measured in an experiment
(inner green box).
(b) NMR pulse sequence for the quantum circuit.
Unfilled rectangles represent $\frac{\pi}{2}$
pulses, while solid rectangles represent $\pi$
pulses.  Tuning of the interaction between qubits is controlled
by varying the $\tau_i$ time period and the
$z$-pulse rotation angle (grey rectangular
shape).
Pulse phases are written above each pulse,
with a bar indicating negative phase.  The
$\tau$ evolution period was fixed at
$\frac{1}{4J_{CH}}$, where $J_{{\rm CH}}$ is the strength of
the scalar coupling.} 
\label{ckt} 
\end{figure}

The quantum circuit to implement the two-qubit entanglement
detection protocol is shown in Fig.\ref{ckt}(a). 
The first block in the circuit (enclosed in a
dashed red box) transforms the $\vert 0 0
\rangle$ pseudopure state to an entangled state with a
desired amount of entanglement. Control of the entanglement
present in the state was achieved by controlling the time
evolution under the non-local interaction Hamiltonian.  A
CNOT gate that achieves this control is represented by a
dashed line. The next block of the circuit (enclosed in a
dashed red box), maps any one of the observables $\langle
O_i\rangle$ ($1\leq i\leq 15$) to the local $z$-magnetization
of one of the qubits, with
$U_{1}^i$ and $U_2^j$ representing local unitaries (as
represented in Table~I). The
dashed green box 
represents the measurement. Only one measurement is
performed in a single experiment.

The NMR pulse sequence to implement the quantum circuit for
entanglement detection using random local measurements,
starting from the pseudopure state $\vert 00 \rangle$  is
shown in Fig.\ref{ckt}(b).  Unfilled rectangles represent
$\frac{\pi}{2}$ pulses while solid rectangles denote
$\pi$ pulses.  Refocusing pulses were used in the middle of
all $J$-evolution periods to compensate for undesired
chemical shift evolution.  Composite pulses  are represented
by $z$  in the pulse sequence, where each composite-$z$
rotation is a sandwich of three pulses:~$xy\bar{x}$.  The
CNOT gate represented by the dashed line in
Fig.\ref{ckt}(a), was achieved experimentally by
controlling the evolution time $\tau_i$, and the angle of
$z$-rotation (the grey-shaded rectangle); $\phi_1^i$
and $\phi_2^j$ are local rotations, and
depend upon which $\langle O_i\rangle$ value is being
measured, and the $\tau$ time interval was set to $
\tau=\frac{1}{2 J_{CH}}$. 
\begin{table}
\caption{\label{table1}
All fifteen observables for two qubits, mapped
to the local \textit{z}-magnetization of  one of the
qubits. This mapping allows a simpler method to measure the
expectation values of the operators $O_j$ and is
completely equivalent to the measurement of the
original local operators.}
\begin{ruledtabular}
\begin{tabular}{c c c c}
\textrm{Observable}&
\textrm{Initial State Mapped to}\\
\colrule
 $\langle O_{1} \rangle$ = Tr[$\rho_{1}.I_{2z}$] & $\rho_1={\rm CNOT}.Y_2.Y_1.\rho_0.Y_1^{\dagger}.Y_2^{\dagger}.{\rm CNOT}^{\dagger}$  \\ 

$\langle O_{2} \rangle$ = Tr[$\rho_{2}.I_{2z}$] & $\rho_2={\rm CNOT}.\bar{X}_2.\bar{X}_1.\rho_0.\bar{X}_1^{\dagger}.\bar{X}_2^{\dagger}.{\rm CNOT}^{\dagger}$  \\ 

$\langle O_{3} \rangle$ = Tr[$\rho_{3}.I_{2z}$] & $\rho_{3}={\rm CNOT}.\rho_0.{\rm CNOT}^{\dagger}$ \\ 

$\langle O_{4} \rangle$ = Tr[$\rho_{4}.I_{2z}$] & $\rho_4={\rm CNOT}.\bar{X}_2.Y_1.\rho_0.Y_1^{\dagger}.\bar{X}_2^{\dagger}.{\rm CNOT}^{\dagger}$  \\ 

$\langle O_{5} \rangle$ = Tr[$\rho_{5}.I_{2z}$] & $\rho_5={\rm CNOT}.Y_1.\rho_0.Y_1^{\dagger}.{\rm CNOT}^{\dagger}$ \\ 

$\langle O_{6} \rangle$ = Tr[$\rho_{6}.I_{2z}$] & $\rho_6={\rm CNOT}.\bar{Y}_2.X_1.\rho_0.X_1^{\dagger}.\bar{Y}_2^{\dagger}.{\rm CNOT}^{\dagger}$ \\ 

$\langle O_{7} \rangle$ = Tr[$\rho_{7}.I_{2z}$] & $\rho_{7}={\rm CNOT}.X_1.\rho_0.X_1^{\dagger}.{\rm CNOT}^{\dagger}$  \\ 

$\langle O_{8} \rangle$ = Tr[$\rho_{8}.I_{2z}$] & $\rho_{8}={\rm CNOT}.\bar{Y}_2.\rho_0.\bar{Y}_2^{\dagger}.{\rm CNOT}^{\dagger}$  \\ 

$\langle O_{9} \rangle$ = Tr[$\rho_{9}.I_{2z}$] & $\rho_{9}={\rm CNOT}.X_2.\rho_0.X_2^{\dagger}.{\rm CNOT}^{\dagger}$ \\ 

$\langle O_{10} \rangle$ = Tr[$\rho_{10}.I_{1z}$] & $\rho_{10}=\bar{Y}_1.\rho_0.\bar{Y}_1^{\dagger}$ \\ 

$\langle O_{11} \rangle$ = Tr[$\rho_{11}.I_{1z}$] & $\rho_{11}=X_1.\rho_0.X_1^{\dagger}$ \\ 

$\langle O_{12} \rangle$ = Tr[$\rho_0.I_{1z}$] & $\rho_0$ is initial state \\ 

$\langle O_{13} \rangle$ = Tr[$\rho_{13}.I_{2z}$] & $\rho_{13}=\bar{Y}_2.\rho_0.\bar{Y}_2^{\dagger}$  \\ 

$\langle O_{14} \rangle$ = Tr[$\rho_{14}.I_{2z}$] & $\rho_{14}=X_2.\rho_0.X_2^{\dagger}$ \\ 

$\langle O_{15} \rangle$ = Tr[$\rho_{0}.I_{2z}$] & $\rho_0$ is initial state  \\ 
\end{tabular}
\end{ruledtabular}
\end{table}

\begin{figure}[h]
\includegraphics[angle=0,scale=1]{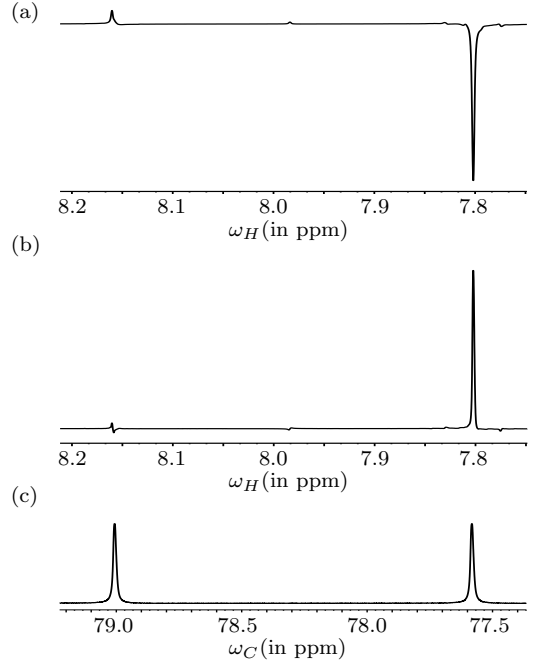} 
\caption{NMR spectra of ${}^{1}$H and ${}^{13}$C nuclei, 
showing the experimentally measured expectation values of
(a) $\langle O_1 \rangle$, (b) $\langle O_2 \rangle$, and
(c) $\langle O_3 \rangle$ respectively, of the
Bell state $\vert \phi^{-}\rangle = \frac{1}{2}(\vert 00
\rangle - \vert 11 \rangle)$. The expectation values have
been measured by the NMR pulse sequences given in Table~I.
}
\label{oisfig} 
\end{figure}

We begin by describing an explicit
example to demonstrate how  the  SDP
can be used
to construct an entanglement witness.
Consider the Bell state $\vert
\phi^{-} \rangle=\frac{1}{\sqrt{2}}
(\vert
00\rangle - \vert 11\rangle)$.
The corresponding
density matrix can be written as a linear superposition of
product operators:
\begin{equation}
\rho = \frac{I}{4}+a 2 I_{1x}I_{2x}+b 2 I_{1y}I_{2y}+c 2 I_{1z}I_{2z}
\end{equation}
where $b=c=-a=\frac{1}{2}$.
Since it is known that the given state is
entangled, the corresponding
entanglement witness can be constructed
as~\cite{guhne-pr-09}:
\begin{eqnarray}
W_{\phi^-} &=& c_{opt}I-\rho \nonumber \\
&=& \frac{I}{4}-a 2 I_{1x}I_{2x}-b 2 I_{1y}I_{2y}-c 2 I_{1z}I_{2z}
\end{eqnarray}
where $c_{opt}$ is the smallest possible value such that the
witness is positive on all separable states;
for Bell states $c_{opt}$ is $\frac{1}{2}$.  Noting
that $\mbox{Tr}(\rho W_{\phi^-})=-\frac{1}{2}<0$, hence by
definition $W_{\phi^-}$ detects the presence of entanglement
in $\rho$.  However, our detection protocol has to deal with
the situation when the state is unknown.  
The question now arises whether the SDP method is able to
find the minimum value of ${\bf c.m}$ such that the correct
$W_{\phi^-}$ is constructed?
For the Bell state
$\vert \phi^{-} \rangle$, the expectation values for
$\langle O_1 \rangle$, $\langle O_2 \rangle$ and $\langle
O_3 \rangle$ yield $-\frac{1}{2}$, $\frac{1}{2}$ and
$\frac{1}{2}$ respectively.  The experimental NMR spectra
obtained after measuring $\langle O_1 \rangle$, $\langle O_2
\rangle$ and $\langle O_3 \rangle$ are shown in
Fig.\ref{oisfig}, with measured expectation values of
$-0.490 \pm 0.021, 0.487 \pm 0.030$ and 
$0.479 \pm 0.015$ respectively (these values
correspond to the area under the absorptive peaks normalized
with respect to the pseudopure state).  These experimental
expectation values are used to construct the vector
\textbf{m}.  The SDP protocol performs minimization
under the given constraints and for this Bell state, is indeed
able to construct $W_{\phi^-}$ as
well as the exact values of $a$, $b$ and $c$ which make up the
vector \textbf{c}.  
Since the  minimum value of
$\textbf{c.m}<0$ is achieved, it confirms the presence of
entanglement in the state.
\begin{table}
\caption{\label{table2} Results of
entanglement detection via local measurements followed
by SDP. States are labeled as
$B$, $S$ or $E$ indicating 
maximally entangled, separable or non-maximally entangled
respectively. The second and third columns contain the 
theoretically expected
and experimentally obtained values of 
the entanglement parameter negativity ($\cal{N}$). 
The $\surd$ in the last column
indicates the success of the experimental protocol in
detecting entanglement.}
\begin{ruledtabular}
\begin{tabular}{c|c c c}
\textrm{State}&\multicolumn{2}{c}{$\cal{N}$} & \textrm{Entanglement}\\
& \textrm{Theory} & \textrm{Exp} & \textrm{Detected}\\
 \colrule
$B_1$ & 0.500 & 0.486 $\pm$ 0.011 & $\surd$ \\
$B_2$ & 0.500 & 0.480 $\pm$ 0.013 & $\surd$ \\
$B_3$ & 0.500 & 0.471 $\pm$ 0.021 & $\surd$ \\
$B_4$ & 0.500 & 0.466 $\pm$ 0.025 & $\surd$ \\
$S_1$ & 0.000 & 0.000 $\pm$ 0.000 & $\surd$ \\
$S_2$ & 0.000 & 0.000 $\pm$ 0.000 & $\surd$ \\
$E_1$ & 0.052 & 0.081 $\pm$ 0.005 & $\times$ \\
$E_2$ & 0.104 & 0.088 $\pm$ 0.024 & $\times$ \\
$E_3$ & 0.155 & 0.177 $\pm$ 0.015 & $\surd$ \\
$E_4$ & 0.203 & 0.182 $\pm$ 0.031 & $\surd$ \\
$E_5$ & 0.250 & 0.212 $\pm$ 0.029 & $\surd$ \\
$E_6$ & 0.294 & 0.255 $\pm$ 0.033 & $\surd$ \\
$E_7$ & 0.335 & 0.297 $\pm$ 0.045 & $\surd$ \\
$E_8$ & 0.372 & 0.351 $\pm$ 0.039 & $\surd$ \\
$E_9$ & 0.405 & 0.400 $\pm$ 0.033 & $\surd$ \\
$E_{10}$ & 0.433 & 0.410 $\pm$ 0.040 & $\surd$ \\
$E_{11}$ & 0.457 & 0.430 $\pm$ 0.037 & $\surd$ \\
$E_{12}$ & 0.476 & 0.444 $\pm$ 0.029 & $\surd$ \\
$E_{13}$ & 0.489 & 0.462 $\pm$ 0.022 & $\surd$ \\
$E_{14}$ & 0.497 & 0.473 $\pm$ 0.025 & $\surd$ \\
\end{tabular}
\end{ruledtabular}
\end{table}

\begin{figure}
\includegraphics[angle=0,scale=1.0]{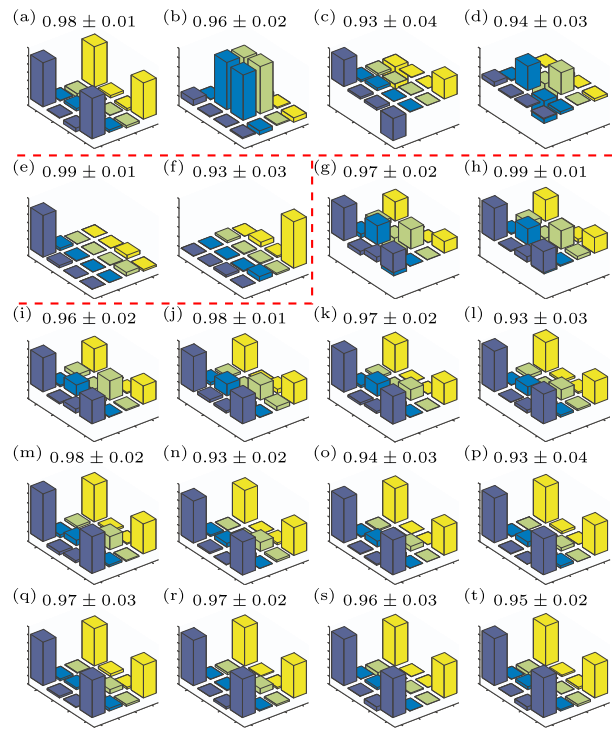}
\caption{(Color online) Real part of the 
tomographed density matrix for
the states described in Table~II.
Maximally entangled Bell
states are shown in (a)-(d), while 
separable states are shown in (e)-(f). 
The tomographs (g)-(t)
represent states of different degrees of
entanglement.  
The state fidelity is written 
above each tomograph.
The 
arrangement of the rows and
columns of the bar graphs is as per the
computational basis of the two-qubit system $\{
\vert 00 \rangle, \vert 01 \rangle, \vert 10
\rangle, \vert 11 \rangle\}$.
}
\label{tomofig}
\end{figure}

We now turn to the detection of entanglement in
states with varying amounts of entanglement.
We experimentally implemented the entanglement
detection protocol on several different states:~four
maximally entangled states (labeled as $B_1$, $B_2$ etc), two
separable states (labeled as $S_1$ and $S_2$) and fourteen
non-maximally entangled states (labeled as $E_1$, $E_2$,
$E_3$...).
To prepare the fourteen
entangled states $E_1$ to $E_{14}$ (having
different amounts of entanglement), the control on
the amount of entanglement in the state was
achieved by varying the time interval $\tau_i$ and
the angle ($\theta$) of the $z$-rotation
(Fig.\ref{ckt}(b)).  We used
$\theta=n\frac{\pi}{30}$ and
$\tau_i=n\frac{1}{30J}$ with $1\leq n\leq 14$.
These choices for $\theta$ and $\tau_i$ represent
a variation of the rotation angle in a two-qubit
controlled-rotation NMR gate and led to a wide
range of entanglement in the generated states 
(as tabulated in Table~II). 
To characterize the amount of entanglement, we used the 
negativity
$\cal{N}$~\cite{horodecki-pla-96} 
as entanglement parameter: 
\begin{equation}
{\cal N} = \| \rho^{PT} {\|} - 1
\end{equation}
where $\rho^{PT}$ denotes partial transpose with respect
to one of the qubits,  
and $\|  \|$ 
represent the trace norm.
A non-zero negativity 
confirms the presence of entanglement 
in $\rho$ and can be used as a quantitative
measure of entanglement.   
The states prepared
ranged from nearly separable ($E_1$, $E_2$ with a
low value of negativity) to nearly maximally
entangled ($E_{13}$, $E_{14}$ with high negativity
values). 
The experimental results of the
entanglement detection protocol for two qubits are tabulated
in Table~II.  
For some of the non-maximally entangled states, 
more than three local measurements had to be used to
detect entanglement.
For instance, SDP required six local
measurement to
build the vector $\bf{m}$ for the $E_8$ state in Table~II, and to
establish that
{\rm Minimum}($\bf{c.m}$) $<0$.
As is evident from Table~II, this method of
making random local measurements on an unknown state
followed by SDP to construct an entanglement witness, is
able to successfully detect the presence of quantum
entanglement in almost all the experimentally created
states.  The protocol failed to detect entanglement in the
states $E_1$ and $E_2$, a possible reason for this being that
these states have a very low negativity value (very
little entanglement), which  is of the order of the
experimental error.  
In order to validate our experimental results 
we also performed quantum state tomography of all
experimentally prepared states.
The resulting
tomographs and respective fidelities are shown in
Fig.\ref{tomofig} and the negativity parameter
obtained from the experimental tomographs in each
case is tabulated in Table~II.
Fig.\ref{tomofig}(a)-(d) correspond to the maximally
entangled Bell states $B_1$ to $B_{4}$
respectively, while Fig.\ref{tomofig}(e) and (f) are tomographs
for the separable states $S_1$ and $S_2$
respectively, and Fig.\ref{tomofig}(g)-(t) correspond
to the states $E_1$ to $E_{14}$ respectively.  
The fidelity of each
experimentally prepared state is given above its
tomograph in the figure. Only the real parts of the
experimental tomographs are shown, as the imaginary parts
of the experimental tomographs turned out to be negligible.
\section{Detecting entanglement in bipartite
$2 \otimes 3$ dimensional systems}
\label{hybrid}
The orthonormal basis states for a $2 \otimes 3$-dimensional 
qubit-qutrit system
$\{\vert i j \rangle~:~i=0,1, j=0,1,2 \}$ 
can be written in 
the computational basis for the qubit ($\{ \vert 0\rangle, 
\vert 1\rangle \}$)
and the qutrit ($ \{ \vert 0\rangle, \vert 1\rangle,
\vert 2\rangle \}$)
respectively.  
It has been previously shown that 
any arbitrary pure state of a hybrid qubit-qutrit 
$2 \otimes 3$ system can be
transformed to one of the states of
a two-parameter class (with two real
parameters), via
local operations and classical
communication (LOCC), and that states
in this class are invariant under
unitary operations of the form $U \otimes U$
on the $2 \otimes 3$ system~\cite{chi-jpamg-03}.
The state for such a bipartite $2\otimes 3$ dimensional
system can be written as~\cite{chi-jpamg-03}
\begin{eqnarray}
\rho &=& \alpha \left[\vert 02\rangle \langle 02 
\vert + \vert 12\rangle \langle 12 \vert \right]
+\beta [ \vert \phi^+\rangle \langle \phi^+\vert + \vert \phi^
-\rangle \langle \phi^-\vert
\nonumber \\
&& + \vert \psi^+\rangle \langle \psi^+ \vert  ]+
\gamma \vert \psi^-\rangle \langle \psi^- \vert
\label{tpr}
\end{eqnarray}
where
\begin{equation}
|\phi^{\pm}\rangle=\frac{1}{\sqrt{2}}\left(|00\rangle \pm |11\rangle\right)
\, {\rm and} \,\,\, 
|\psi^{\pm}\rangle=\frac{1}{\sqrt{2}}\left(|01\rangle \pm |10\rangle\right)
\label{bell12}
\end{equation}
are the maximally entangled Bell states.

The requirement of unit trace
places a constraint on 
the real parameters $\alpha, \beta$ and $\gamma$ 
\begin{equation}
2\alpha+3\beta+\gamma=1
\end{equation}
This constraint implies that one can eliminate 
one of the three parameters and we
have chosen to rewrite $\beta$ in terms of
$\alpha$ and $\gamma$; however the entire analysis
is valid for the other choices as well.
The domains for $\alpha, \gamma$ can be
calculated  from the unit trace condition and turn
out to be $ 0 \le \alpha \le 1/2$ and $0 \le \gamma \le 1$.

The Peres-Horodecki positive partial transposition (PPT) 
criterion is a necessary and sufficient condition
for $2\otimes3$ dimensional 
systems, and can hence be 
used to characterize the entanglement of
$\rho$ via the entanglement parameter
termed negativity ($\cal{N}$).
The partial transpose with respect to the qubit, 
for the two-parameter class
of states defined in Eqn.(\ref{tpr}) can be
written as
\begin{eqnarray} 
\rho^{PT} &=& \alpha \left[\vert 02\rangle
\langle 02 \vert + \vert 12\rangle \langle 12
\vert \right]+\frac{(\beta+\gamma)}{2} [\vert 01\rangle \langle 01
\vert \nonumber \\ && + \vert 10\rangle \langle 10 \vert+
\vert \phi^-\rangle \langle \phi^- \vert ] +
\frac{(3\beta-\gamma)}{2}\vert \phi^+\rangle \langle \phi^+
\vert 
\end{eqnarray}

The negativity ${\cal N}(\rho)$ 
for the two-parameter class of states 
can be calculated from its partial transpose and
is given by~\cite{chi-jpamg-03}:
\begin{equation}
{\cal N}(\rho)=max\lbrace (2\alpha +2\gamma-1), 0\rbrace
\end{equation}
Clearly, 
states with $1/2 < \alpha+\gamma \leq 1$ have non-zero
negativity 
(i.e. are NPT) and are hence entangled.

We extend the Bloch representation for qubits to a 
qubit-qutrit system described by a
$2\otimes3$ dimensional hybrid linear vector space. An
operator $O$ operating on this 
combined space can be written as~\cite{jami-ind-05}
\begin{eqnarray}
O &=& \frac{1}{6} [ I_2 \otimes I_3 +
\sigma^A . \vec{u} \otimes I_3 
+\sqrt{3} I_2 \otimes \lambda^B . \vec{v}
\nonumber \\
&& + \sum_{i=1}^{3} \sum_{j=1}^{8} \beta_{ij}
(\sigma^{A}{_i} \otimes \lambda^{B}_{j})
]
\label{decomposeeqn}
\end{eqnarray}
where 
$\vec{u}$ and $\vec{v}$ are vectors belonging to a linear
vector space of dimension 3 and 8 respectively, $I_2$ and
$I_3$ are identity matrices of dimensions 2 and 3,
$\sigma_i$ are the
Pauli spin matrices used to span operators acting on the
Hilbert space of the qubit, and $\lambda_j$ are the Gell-Mann
matrices~\cite{gellmann-book}, used to span operators acting
on the Hilbert space of the qutrit; other isomorphic choices are
equally valid.

A Hermitian witness operator can be constructed for every entangled
quantum state and the expectation value of the witness
operator can be locally measured by decomposing the operator
as a weighted sum of projectors onto product state
vectors~\cite{horodecki-pla-96,bourennane-prl-04,brandao-pra-05}.
The $\rho$ for the $2 \otimes 3$ system 
given in Eqn.(\ref{tpr}) is NPT for
$0.5 < (\alpha + \gamma) \leq 1$. The eigenvalues for
$\rho^{PT}$ (where $PT$ represents partial transposition
with respect to the qubit) are: $\alpha$,
$\frac{1}{2}(1-2\alpha-2\gamma)$ and
$\frac{1}{6}(1-2\alpha+2\gamma)$.   The eigenvalue
$\frac{1}{2}(1-2\alpha-2\gamma)$  remains  negative for NPT
states and we denote its
corresponding eigenvector by $ \vert \eta \rangle$. 
The corresponding entanglement witness operator can be
written as
$W= (\vert \eta \rangle \langle \eta \vert)^{PT}$ with
its matrix representation
\begin{equation}
W = \left( 
\begin{array}{cccccc}
\frac{1}{2}&0&0&0&0&0\\
0&0&0&\frac{1}{2}&0&0\\
0&0&0&0&0&0\\
0&\frac{1}{2}&0&0&0&0\\
0&0&0&0&0&0\\ 
0&0&0&0&0&\frac{1}{2}\\
\end{array} 
\right)
\label{wmatrix}
\end{equation}
The entanglement witness $W$  
is capable of detecting entanglement of
the $2 \otimes 3$ dimensional $\rho$ given in 
Eqn.(\ref{tpr}).

We now turn to the decomposition of the entanglement witness
$W$ in terms of local observables, so that we can use it
to detect entanglement of the two-parameter class of states
of the $2 \otimes 3$ dimensional $\rho$.
The explicit decomposition of $W$ 
as per Eqn.(\ref{decomposeeqn}) results in the following:
\begin{equation}
\vec{u}=\left[ \begin{array}{c}
0\\
0\\
0
\end{array} \right]
,\,\,\, 
\vec{v}=\left[ \begin{array}{c}
0\\
0\\
0\\
0\\
0\\
0\\
0\\
1
\end{array} \right]
,\,\,\, 
\beta=\left[ \begin{array}{cccccccc}
\frac{1}{2}&0&0&0&0&0&0&0\\
0&\frac{1}{2}&0&0&0&0&0&0\\
0&0&\frac{1}{2}&0&0&0&0&0
\end{array} \right]
\label{blochdec}
\end{equation}
The components of  $\vec{u}$ and $\vec{v}$ i.e. $u_i$ 
($ i=1,2,3 $) and $v_j$ ($ j=1,2,3,..8 $) were
obtained from $u_i={\rm Tr}[W (\sigma_i\otimes I_3)]$ and
$v_j={\rm Tr}[W (I_2\otimes\lambda_j)]$. Similarly the elements
of the matrix $\beta$ can be obtained from
$\beta_{ij}={\rm Tr}[W (\sigma_i\otimes\lambda_j)]$.
There are thirty five real coefficients in the expansion in
Eqn.(\ref{decomposeeqn}), of which three coefficients
constitute $\vec{u}$, eight coefficients constitute
$\vec{v}$ and the remaining twenty four are contained  in
the $\beta$ matrix. Each non-zero entry in $\vec{u}$,
$\vec{v}$ or the $\beta$ matrix is the contribution of the
corresponding qubit-qutrit product
operator~\cite{ernst-book} used in the construction of
operator $W$. Hence one can infer
by inspection of the  non-zero matrix entries in
Eqn.(\ref{blochdec}), that one requires the expectation
values of at least four operators in a given state, in order
to experimentally construct the witness operator $W$.

\begin{figure}[t]
\includegraphics[scale=1]{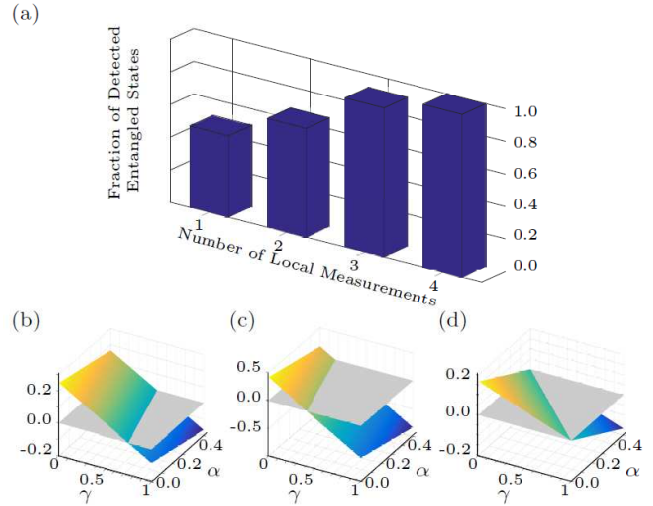}
\caption{(Color online) (a) Bar graph showing the
fraction of the detected entangled states plotted
as a function of the number of local measurements,
from the simulation on the qubit-qutrit system.
Panels (b), (c), (d) show plots of ${\rm
Tr}(W_{I}\rho_{AB})$, ${\rm Tr}(W_{II}\rho_{AB})$,
and ${\rm Tr}(W_{III}\rho_{AB})$ respectively, for
$0\le\alpha\le 0.5$ and $0\le\gamma\le 1$.  The
entanglement witness operators $W_{I}, W_{II},
W_{III}$ are constructed by choosing sets of one,
two or three random local measurements at a time,
respectively.  A reference plane (gray-shaded) at
vanishing trace is also plotted to better
differentiate positive and negative values.  }
\label{dffig} \end{figure}

While the maximum number of expectation values
required to be measured is four, the question
remains if this is an optimal set or can we find
a smaller set which will still be able to detect
entanglement.  We hence computed the fraction of
entanglement detected, by gradually increasing the
number of local observations, and the results of
the simulation are depicted in Fig.\ref{dffig}(a)
as a bar chart.  We note here that even if we
measure only one observable (one element of the
$\beta$ matrix in Eqn.(\ref{blochdec})), half of
the randomly generated entangled states are
detected.  As the number of measured observables
is increased, the fraction of detected entangled
states improves, as shown in Fig.\ref{dffig}(a).
To generate the bar plots in Fig.\ref{dffig}, we
began by selecting only one random local
measurement out of the maximum thirty five
possible measurements.  Only those choices are
valid which will establish a decomposable
entanglement witness of unit trace.  For one such
choice (denoted by $W_I$), the ${\rm Tr}(W_I\rho)$
is plotted in Fig.\ref{dffig}(b) in the range
$0\le\alpha\le 0.5$ and $0\le\gamma\le 1$. As is
evident, this $W_I$ (based on only one random
local measurement) does not detect all the
entangled states which were detected by $W$.  The
fraction of entangled states detected by $W_I$ can
be computed from geometry i.e. how much area that
is spanned by the parameters $\alpha$ and $\gamma$
represents entangled states and how much of that
area is  detected by the corresponding
entanglement witness operator.  When we consider
random local measurements chosen two at a time, to
construct a valid entanglement witness ( $W_{II}$
in Fig.\ref{dffig}(c)), the detected fraction of
entangled states improves from 0.50 to 0.67 (the
second bar in Fig.\ref{dffig}(a)).  One can
observe from the geometry that $W_{II}$ detects
more entangled states as compared to $W_I$, but
this fraction is still smaller than those detected
by $W$.  The result of choosing three random local
measurements (denoted by witness operator
$W_{III}$) is plotted in Fig.\ref{dffig}(d), and
this detects 83.3\% of the total entangled states
(the third bar in Fig.\ref{dffig}(a)).
Increasing the set of random local measurements
hence increases the probability of detecting
entanglement.  The worst case detection fraction
is shown in Fig.\ref{dffig}(a), when choosing
random local measurements.  A fraction of 1 in the
Fig.\ref{dffig}(a) implies that the corresponding
set of four random local measurements, will always
be able to detect entanglement in the state, if it
exists.
\section{Concluding Remarks} \label{concl} We have
experimentally demonstrated how a scheme based on
local random measurements detects the presence of
entanglement, on a two-qubit NMR quantum
information processor.  An optimal set of random
local measurements was arrived at via
semi-definite programming to construct
entanglement witnesses that detect bipartite
entanglement.  The local measurements on each
qubit were converted into a single measurement on
one of the qubits by transforming the state. This
was done to simplify the experimental scheme and
is completely equivalent to the originally
intended local measurements.

We have extended the scheme based on random local
measurements to hybrid systems, where qudits of
different dimensionality are involved.  For the
particular case of a qubit-qutrit system, we
perform a simulation to demonstrate the optimality
of the detection scheme.  Characterization of
entangled states of qudits is a daunting task, and
this work holds promise for further research in
this direction.  Efforts are on to extend the
scheme to experimentally detect the presence of
multipartite entanglement in systems of three and
more coupled qubits.
\acknowledgments
All experiments were performed on a Bruker
Avance-III 600 MHz FT-NMR spectrometer at the NMR
Research Facility at IISER Mohali.  Arvind
acknowledges funding from DST India under Grant
number EMR/2014/000297.  KD acknowledges funding
from DST India under Grant number EMR/2015/000556.


\begin{thebibliography}{43}%
\makeatletter
\providecommand \@ifxundefined [1]{%
 \@ifx{#1\undefined}
}%
\providecommand \@ifnum [1]{%
 \ifnum #1\expandafter \@firstoftwo
 \else \expandafter \@secondoftwo
 \fi
}%
\providecommand \@ifx [1]{%
 \ifx #1\expandafter \@firstoftwo
 \else \expandafter \@secondoftwo
 \fi
}%
\providecommand \natexlab [1]{#1}%
\providecommand \enquote  [1]{``#1''}%
\providecommand \bibnamefont  [1]{#1}%
\providecommand \bibfnamefont [1]{#1}%
\providecommand \citenamefont [1]{#1}%
\providecommand \href@noop [0]{\@secondoftwo}%
\providecommand \href [0]{\begingroup \@sanitize@url \@href}%
\providecommand \@href[1]{\@@startlink{#1}\@@href}%
\providecommand \@@href[1]{\endgroup#1\@@endlink}%
\providecommand \@sanitize@url [0]{\catcode `\\12\catcode `\$12\catcode
  `\&12\catcode `\#12\catcode `\^12\catcode `\_12\catcode `\%12\relax}%
\providecommand \@@startlink[1]{}%
\providecommand \@@endlink[0]{}%
\providecommand \url  [0]{\begingroup\@sanitize@url \@url }%
\providecommand \@url [1]{\endgroup\@href {#1}{\urlprefix }}%
\providecommand \urlprefix  [0]{URL }%
\providecommand \Eprint [0]{\href }%
\@ifxundefined \urlstyle {%
  \providecommand \doi  [0]{\begingroup \@sanitize@url \@doi}%
  \providecommand \@doi [1]{\endgroup \@@startlink {\doibase
  #1}doi:\discretionary {}{}{}#1\@@endlink }%
}{%
  \providecommand \doi  [0]{doi:\discretionary{}{}{}\begingroup
  \urlstyle{rm}\Url }%
}%
\providecommand \doibase [0]{http://dx.doi.org/}%
\providecommand \Doi [0]{\begingroup \@sanitize@url \@Doi }%
\providecommand \@Doi  [1]{\endgroup\@@startlink{\doibase#1}\@@Doi}%
\providecommand \@@Doi [1]{#1\@@endlink}%
\providecommand \selectlanguage [0]{\@gobble}%
\providecommand \bibinfo  [0]{\@secondoftwo}%
\providecommand \bibfield  [0]{\@secondoftwo}%
\providecommand \translation [1]{[#1]}%
\providecommand \BibitemOpen [0]{}%
\providecommand \bibitemStop [0]{}%
\providecommand \bibitemNoStop [0]{.\EOS\space}%
\providecommand \EOS [0]{\spacefactor3000\relax}%
\providecommand \BibitemShut  [1]{\csname bibitem#1\endcsname}%
\bibitem [{\citenamefont {Horodecki}\ \emph {et~al.}(2009)\citenamefont
  {Horodecki}, \citenamefont {Horodecki}, \citenamefont {Horodecki},\ and\
  \citenamefont {Horodecki}}]{horodecki-rmp-09}%
  \BibitemOpen
  \bibfield  {author} {\bibinfo {author} {\bibfnamefont {R.}~\bibnamefont
  {Horodecki}}, \bibinfo {author} {\bibfnamefont {P.}~\bibnamefont
  {Horodecki}}, \bibinfo {author} {\bibfnamefont {M.}~\bibnamefont
  {Horodecki}}, \ and\ \bibinfo {author} {\bibfnamefont {K.}~\bibnamefont
  {Horodecki}},\ }\href@noop {} {\bibfield  {journal} {\bibinfo  {journal}
  {Rev. Mod. Phys.},\ }\textbf {\bibinfo {volume} {81}},\ \bibinfo {pages}
  {865} (\bibinfo {year} {2009})}\BibitemShut {NoStop}%
\bibitem [{\citenamefont {Huber}\ \emph {et~al.}(2011)\citenamefont {Huber},
  \citenamefont {Schimpf}, \citenamefont {Gabriel}, \citenamefont {Spengler},
  \citenamefont {Bruss},\ and\ \citenamefont {Hiesmayr}}]{huber-pra-11}%
  \BibitemOpen
  \bibfield  {author} {\bibinfo {author} {\bibfnamefont {M.}~\bibnamefont
  {Huber}}, \bibinfo {author} {\bibfnamefont {H.}~\bibnamefont {Schimpf}},
  \bibinfo {author} {\bibfnamefont {A.}~\bibnamefont {Gabriel}}, \bibinfo
  {author} {\bibfnamefont {C.}~\bibnamefont {Spengler}}, \bibinfo {author}
  {\bibfnamefont {D.}~\bibnamefont {Bruss}}, \ and\ \bibinfo {author}
  {\bibfnamefont {B.~C.}\ \bibnamefont {Hiesmayr}},\ }\href@noop {} {\bibfield
  {journal} {\bibinfo  {journal} {Phys. Rev. A},\ }\textbf {\bibinfo {volume}
  {83}},\ \bibinfo {pages} {022328} (\bibinfo {year} {2011})}\BibitemShut
  {NoStop}%
\bibitem [{\citenamefont {Jungnitsch}\ \emph {et~al.}(2011)\citenamefont
  {Jungnitsch}, \citenamefont {Moroder},\ and\ \citenamefont
  {Guhne}}]{jungnitsch-prl-11}%
  \BibitemOpen
  \bibfield  {author} {\bibinfo {author} {\bibfnamefont {B.}~\bibnamefont
  {Jungnitsch}}, \bibinfo {author} {\bibfnamefont {T.}~\bibnamefont {Moroder}},
  \ and\ \bibinfo {author} {\bibfnamefont {O.}~\bibnamefont {Guhne}},\
  }\href@noop {} {\bibfield  {journal} {\bibinfo  {journal} {Phys. Rev.
  Lett.},\ }\textbf {\bibinfo {volume} {106}},\ \bibinfo {pages} {190502}
  (\bibinfo {year} {2011})}\BibitemShut {NoStop}%
\bibitem [{\citenamefont {Wallman}\ and\ \citenamefont
  {Bartlett}(2012)}]{wallman-pra-12}%
  \BibitemOpen
  \bibfield  {author} {\bibinfo {author} {\bibfnamefont {J.~J.}\ \bibnamefont
  {Wallman}}\ and\ \bibinfo {author} {\bibfnamefont {S.~D.}\ \bibnamefont
  {Bartlett}},\ }\href@noop {} {\bibfield  {journal} {\bibinfo  {journal}
  {Phys. Rev. A},\ }\textbf {\bibinfo {volume} {85}},\ \bibinfo {pages}
  {024101} (\bibinfo {year} {2012})}\BibitemShut {NoStop}%
\bibitem [{\citenamefont {White}\ \emph {et~al.}(1999)\citenamefont {White},
  \citenamefont {James}, \citenamefont {Eberhard},\ and\ \citenamefont
  {Kwiat}}]{white-prl-98}%
  \BibitemOpen
  \bibfield  {author} {\bibinfo {author} {\bibfnamefont {A.~G.}\ \bibnamefont
  {White}}, \bibinfo {author} {\bibfnamefont {D.~F.~V.}\ \bibnamefont {James}},
  \bibinfo {author} {\bibfnamefont {P.~H.}\ \bibnamefont {Eberhard}}, \ and\
  \bibinfo {author} {\bibfnamefont {P.~G.}\ \bibnamefont {Kwiat}},\ }\href@noop
  {} {\bibfield  {journal} {\bibinfo  {journal} {Phys. Rev. Lett.},\ }\textbf
  {\bibinfo {volume} {83}},\ \bibinfo {pages} {3103} (\bibinfo {year}
  {1999})}\BibitemShut {NoStop}%
\bibitem [{\citenamefont {Thew}\ \emph {et~al.}(2002)\citenamefont {Thew},
  \citenamefont {Nemoto}, \citenamefont {White},\ and\ \citenamefont
  {Munro}}]{thew-pra-02}%
  \BibitemOpen
  \bibfield  {author} {\bibinfo {author} {\bibfnamefont {R.~T.}\ \bibnamefont
  {Thew}}, \bibinfo {author} {\bibfnamefont {K.}~\bibnamefont {Nemoto}},
  \bibinfo {author} {\bibfnamefont {A.~G.}\ \bibnamefont {White}}, \ and\
  \bibinfo {author} {\bibfnamefont {W.~J.}\ \bibnamefont {Munro}},\ }\href@noop
  {} {\bibfield  {journal} {\bibinfo  {journal} {Phys. Rev. A},\ }\textbf
  {\bibinfo {volume} {66}},\ \bibinfo {pages} {012303} (\bibinfo {year}
  {2002})}\BibitemShut {NoStop}%
\bibitem [{\citenamefont {Behrman}\ and\ \citenamefont
  {Steck}(2013)}]{behrman-qic-13}%
  \BibitemOpen
  \bibfield  {author} {\bibinfo {author} {\bibfnamefont {E.~C.}\ \bibnamefont
  {Behrman}}\ and\ \bibinfo {author} {\bibfnamefont {J.~E.}\ \bibnamefont
  {Steck}},\ }\href@noop {} {\bibfield  {journal} {\bibinfo  {journal} {Quant.
  Inf. Comp.},\ }\textbf {\bibinfo {volume} {13}},\ \bibinfo {pages} {36}
  (\bibinfo {year} {2013})}\BibitemShut {NoStop}%
\bibitem [{\citenamefont {Spedalieri}(2012)}]{spedalieri-pra-12}%
  \BibitemOpen
  \bibfield  {author} {\bibinfo {author} {\bibfnamefont {F.~M.}\ \bibnamefont
  {Spedalieri}},\ }\href@noop {} {\bibfield  {journal} {\bibinfo  {journal}
  {Phys. Rev. A},\ }\textbf {\bibinfo {volume} {86}},\ \bibinfo {pages}
  {062311} (\bibinfo {year} {2012})}\BibitemShut {NoStop}%
\bibitem [{\citenamefont {Lewenstein}\ \emph {et~al.}(2000)\citenamefont
  {Lewenstein}, \citenamefont {Kraus}, \citenamefont {Cirac},\ and\
  \citenamefont {Horodecki}}]{lewenstein-pra-00}%
  \BibitemOpen
  \bibfield  {author} {\bibinfo {author} {\bibfnamefont {M.}~\bibnamefont
  {Lewenstein}}, \bibinfo {author} {\bibfnamefont {B.}~\bibnamefont {Kraus}},
  \bibinfo {author} {\bibfnamefont {J.~I.}\ \bibnamefont {Cirac}}, \ and\
  \bibinfo {author} {\bibfnamefont {P.}~\bibnamefont {Horodecki}},\ }\href@noop
  {} {\bibfield  {journal} {\bibinfo  {journal} {Phys. Rev. A},\ }\textbf
  {\bibinfo {volume} {62}},\ \bibinfo {pages} {052310} (\bibinfo {year}
  {2000})}\BibitemShut {NoStop}%
\bibitem [{\citenamefont {Guhne}\ \emph {et~al.}(2003)\citenamefont {Guhne},
  \citenamefont {Hyllus}, \citenamefont {Bruss}, \citenamefont {Ekert},
  \citenamefont {Lewenstein}, \citenamefont {Macchiavello},\ and\ \citenamefont
  {Sanpera}}]{guhne-jmo-03}%
  \BibitemOpen
  \bibfield  {author} {\bibinfo {author} {\bibfnamefont {O.}~\bibnamefont
  {Guhne}}, \bibinfo {author} {\bibfnamefont {P.}~\bibnamefont {Hyllus}},
  \bibinfo {author} {\bibfnamefont {D.}~\bibnamefont {Bruss}}, \bibinfo
  {author} {\bibfnamefont {A.}~\bibnamefont {Ekert}}, \bibinfo {author}
  {\bibfnamefont {M.}~\bibnamefont {Lewenstein}}, \bibinfo {author}
  {\bibfnamefont {C.}~\bibnamefont {Macchiavello}}, \ and\ \bibinfo {author}
  {\bibfnamefont {A.}~\bibnamefont {Sanpera}},\ }\href@noop {} {\bibfield
  {journal} {\bibinfo  {journal} {J. Mod. Opt.},\ }\textbf {\bibinfo {volume}
  {50}},\ \bibinfo {pages} {1079} (\bibinfo {year} {2003})}\BibitemShut
  {NoStop}%
\bibitem [{\citenamefont {Arrazola}\ \emph {et~al.}(2012)\citenamefont
  {Arrazola}, \citenamefont {Gittsovich},\ and\ \citenamefont
  {Lutkenhaus}}]{arrazola-pra-12}%
  \BibitemOpen
  \bibfield  {author} {\bibinfo {author} {\bibfnamefont {J.~M.}\ \bibnamefont
  {Arrazola}}, \bibinfo {author} {\bibfnamefont {O.}~\bibnamefont
  {Gittsovich}}, \ and\ \bibinfo {author} {\bibfnamefont {N.}~\bibnamefont
  {Lutkenhaus}},\ }\href@noop {} {\bibfield  {journal} {\bibinfo  {journal}
  {Phys. Rev. A},\ }\textbf {\bibinfo {volume} {85}},\ \bibinfo {pages}
  {062327} (\bibinfo {year} {2012})}\BibitemShut {NoStop}%
\bibitem [{\citenamefont {Novo}\ \emph {et~al.}(2013)\citenamefont {Novo},
  \citenamefont {Moroder},\ and\ \citenamefont {Guhne}}]{novo-pra-13}%
  \BibitemOpen
  \bibfield  {author} {\bibinfo {author} {\bibfnamefont {L.}~\bibnamefont
  {Novo}}, \bibinfo {author} {\bibfnamefont {T.}~\bibnamefont {Moroder}}, \
  and\ \bibinfo {author} {\bibfnamefont {O.}~\bibnamefont {Guhne}},\
  }\href@noop {} {\bibfield  {journal} {\bibinfo  {journal} {Phys. Rev. A},\
  }\textbf {\bibinfo {volume} {88}},\ \bibinfo {pages} {012305} (\bibinfo
  {year} {2013})}\BibitemShut {NoStop}%
\bibitem [{\citenamefont {Bartkiewicz}\ \emph {et~al.}(2015)\citenamefont
  {Bartkiewicz}, \citenamefont {Beran}, \citenamefont {Lemr}, \citenamefont
  {Norek},\ and\ \citenamefont {Miranowicz}}]{bartkiewicz-pra-15-1}%
  \BibitemOpen
  \bibfield  {author} {\bibinfo {author} {\bibfnamefont {K.}~\bibnamefont
  {Bartkiewicz}}, \bibinfo {author} {\bibfnamefont {J.}~\bibnamefont {Beran}},
  \bibinfo {author} {\bibfnamefont {K.}~\bibnamefont {Lemr}}, \bibinfo {author}
  {\bibfnamefont {M.}~\bibnamefont {Norek}}, \ and\ \bibinfo {author}
  {\bibfnamefont {A.}~\bibnamefont {Miranowicz}},\ }\href@noop {} {\bibfield
  {journal} {\bibinfo  {journal} {Phys. Rev. A},\ }\textbf {\bibinfo {volume}
  {91}},\ \bibinfo {pages} {022323} (\bibinfo {year} {2015})}\BibitemShut
  {NoStop}%
\bibitem [{\citenamefont {Zhao}\ \emph {et~al.}(2013)\citenamefont {Zhao},
  \citenamefont {Zhang}, \citenamefont {Li-Jost},\ and\ \citenamefont
  {Fei}}]{zhao-pra-13}%
  \BibitemOpen
  \bibfield  {author} {\bibinfo {author} {\bibfnamefont {M.-J.}\ \bibnamefont
  {Zhao}}, \bibinfo {author} {\bibfnamefont {T.-G.}\ \bibnamefont {Zhang}},
  \bibinfo {author} {\bibfnamefont {X.}~\bibnamefont {Li-Jost}}, \ and\
  \bibinfo {author} {\bibfnamefont {S.-M.}\ \bibnamefont {Fei}},\ }\href@noop
  {} {\bibfield  {journal} {\bibinfo  {journal} {Phys. Rev. A},\ }\textbf
  {\bibinfo {volume} {87}},\ \bibinfo {pages} {012316} (\bibinfo {year}
  {2013})}\BibitemShut {NoStop}%
\bibitem [{\citenamefont {Miranowicz}\ \emph {et~al.}(2014)\citenamefont
  {Miranowicz}, \citenamefont {Bartkiewicz}, \citenamefont {Jr}, \citenamefont
  {Koashi}, \citenamefont {Imoto},\ and\ \citenamefont
  {Nori}}]{miranowicz-pra-14}%
  \BibitemOpen
  \bibfield  {author} {\bibinfo {author} {\bibfnamefont {A.}~\bibnamefont
  {Miranowicz}}, \bibinfo {author} {\bibfnamefont {K.}~\bibnamefont
  {Bartkiewicz}}, \bibinfo {author} {\bibfnamefont {J.~P.}\ \bibnamefont {Jr}},
  \bibinfo {author} {\bibfnamefont {M.}~\bibnamefont {Koashi}}, \bibinfo
  {author} {\bibfnamefont {N.}~\bibnamefont {Imoto}}, \ and\ \bibinfo {author}
  {\bibfnamefont {F.}~\bibnamefont {Nori}},\ }\href@noop {} {\bibfield
  {journal} {\bibinfo  {journal} {Phys. Rev. A},\ }\textbf {\bibinfo {volume}
  {90}},\ \bibinfo {pages} {062123} (\bibinfo {year} {2014})}\BibitemShut
  {NoStop}%
\bibitem [{\citenamefont {Horodecki}\ \emph {et~al.}(1996)\citenamefont
  {Horodecki}, \citenamefont {Horodecki},\ and\ \citenamefont
  {Horodecki}}]{horodecki-pla-96}%
  \BibitemOpen
  \bibfield  {author} {\bibinfo {author} {\bibfnamefont {M.}~\bibnamefont
  {Horodecki}}, \bibinfo {author} {\bibfnamefont {P.}~\bibnamefont
  {Horodecki}}, \ and\ \bibinfo {author} {\bibfnamefont {R.}~\bibnamefont
  {Horodecki}},\ }\href@noop {} {\bibfield  {journal} {\bibinfo  {journal}
  {Phys. Lett. A},\ }\textbf {\bibinfo {volume} {223}},\ \bibinfo {pages} {1}
  (\bibinfo {year} {1996})}\BibitemShut {NoStop}%
\bibitem [{\citenamefont {Peres}(1996)}]{peres-prl-98}%
  \BibitemOpen
  \bibfield  {author} {\bibinfo {author} {\bibfnamefont {A.}~\bibnamefont
  {Peres}},\ }\href@noop {} {\bibfield  {journal} {\bibinfo  {journal} {Phys.
  Rev. Lett.},\ }\textbf {\bibinfo {volume} {77}},\ \bibinfo {pages} {1413}
  (\bibinfo {year} {1996})}\BibitemShut {NoStop}%
\bibitem [{\citenamefont {Guhne}\ and\ \citenamefont
  {Toth}(2009)}]{guhne-pr-09}%
  \BibitemOpen
  \bibfield  {author} {\bibinfo {author} {\bibfnamefont {O.}~\bibnamefont
  {Guhne}}\ and\ \bibinfo {author} {\bibfnamefont {G.}~\bibnamefont {Toth}},\
  }\href@noop {} {\bibfield  {journal} {\bibinfo  {journal} {Phys. Rep.},\
  }\textbf {\bibinfo {volume} {474}},\ \bibinfo {pages} {1} (\bibinfo {year}
  {2009})}\BibitemShut {NoStop}%
\bibitem [{\citenamefont {Filgueiras}\ \emph {et~al.}(2012)\citenamefont
  {Filgueiras}, \citenamefont {Maciel}, \citenamefont {Auccaise}, \citenamefont
  {Vianna}, \citenamefont {Sarthour},\ and\ \citenamefont
  {Oliveira}}]{filgueiras-qip-12}%
  \BibitemOpen
  \bibfield  {author} {\bibinfo {author} {\bibfnamefont {J.~G.}\ \bibnamefont
  {Filgueiras}}, \bibinfo {author} {\bibfnamefont {T.~O.}\ \bibnamefont
  {Maciel}}, \bibinfo {author} {\bibfnamefont {R.~E.}\ \bibnamefont
  {Auccaise}}, \bibinfo {author} {\bibfnamefont {R.~O.}\ \bibnamefont
  {Vianna}}, \bibinfo {author} {\bibfnamefont {R.~S.}\ \bibnamefont
  {Sarthour}}, \ and\ \bibinfo {author} {\bibfnamefont {I.~S.}\ \bibnamefont
  {Oliveira}},\ }\href@noop {} {\bibfield  {journal} {\bibinfo  {journal}
  {Quant. Inf. Process.},\ }\textbf {\bibinfo {volume} {11}},\ \bibinfo {pages}
  {1883} (\bibinfo {year} {2012})}\BibitemShut {NoStop}%
\bibitem [{\citenamefont {Silva}\ \emph {et~al.}(2013)\citenamefont {Silva},
  \citenamefont {Girolami}, \citenamefont {Auccaise}, \citenamefont {Sarthour},
  \citenamefont {Oliveira}, \citenamefont {Bonagamba}, \citenamefont
  {deAzevedo}, \citenamefont {Soares-Pinto},\ and\ \citenamefont
  {Adesso}}]{silva-prl-13}%
  \BibitemOpen
  \bibfield  {author} {\bibinfo {author} {\bibfnamefont {I.~A.}\ \bibnamefont
  {Silva}}, \bibinfo {author} {\bibfnamefont {D.}~\bibnamefont {Girolami}},
  \bibinfo {author} {\bibfnamefont {R.}~\bibnamefont {Auccaise}}, \bibinfo
  {author} {\bibfnamefont {R.~S.}\ \bibnamefont {Sarthour}}, \bibinfo {author}
  {\bibfnamefont {I.~S.}\ \bibnamefont {Oliveira}}, \bibinfo {author}
  {\bibfnamefont {T.~J.}\ \bibnamefont {Bonagamba}}, \bibinfo {author}
  {\bibfnamefont {E.~R.}\ \bibnamefont {deAzevedo}}, \bibinfo {author}
  {\bibfnamefont {D.~O.}\ \bibnamefont {Soares-Pinto}}, \ and\ \bibinfo
  {author} {\bibfnamefont {G.}~\bibnamefont {Adesso}},\ }\href@noop {}
  {\bibfield  {journal} {\bibinfo  {journal} {Phys. Rev. Lett.},\ }\textbf
  {\bibinfo {volume} {110}},\ \bibinfo {pages} {140501} (\bibinfo {year}
  {2013})}\BibitemShut {NoStop}%
\bibitem [{\citenamefont {Kampermann}\ \emph {et~al.}(2010)\citenamefont
  {Kampermann}, \citenamefont {Bruss}, \citenamefont {Peng},\ and\
  \citenamefont {Suter}}]{suter-pra-10}%
  \BibitemOpen
  \bibfield  {author} {\bibinfo {author} {\bibfnamefont {H.}~\bibnamefont
  {Kampermann}}, \bibinfo {author} {\bibfnamefont {D.}~\bibnamefont {Bruss}},
  \bibinfo {author} {\bibfnamefont {X.}~\bibnamefont {Peng}}, \ and\ \bibinfo
  {author} {\bibfnamefont {D.}~\bibnamefont {Suter}},\ }\href@noop {}
  {\bibfield  {journal} {\bibinfo  {journal} {Phys. Rev. A},\ }\textbf
  {\bibinfo {volume} {81}},\ \bibinfo {pages} {040304} (\bibinfo {year}
  {2010})}\BibitemShut {NoStop}%
\bibitem [{\citenamefont {Peng}\ \emph {et~al.}(2010)\citenamefont {Peng},
  \citenamefont {Zhang}, \citenamefont {Du},\ and\ \citenamefont
  {Suter}}]{suter-pra-10-1}%
  \BibitemOpen
  \bibfield  {author} {\bibinfo {author} {\bibfnamefont {X.}~\bibnamefont
  {Peng}}, \bibinfo {author} {\bibfnamefont {J.}~\bibnamefont {Zhang}},
  \bibinfo {author} {\bibfnamefont {J.}~\bibnamefont {Du}}, \ and\ \bibinfo
  {author} {\bibfnamefont {D.}~\bibnamefont {Suter}},\ }\href@noop {}
  {\bibfield  {journal} {\bibinfo  {journal} {Phys. Rev. A},\ }\textbf
  {\bibinfo {volume} {81}},\ \bibinfo {pages} {042327} (\bibinfo {year}
  {2010})}\BibitemShut {NoStop}%
\bibitem [{\citenamefont {Dogra}\ \emph {et~al.}(2015)\citenamefont {Dogra},
  \citenamefont {Dorai},\ and\ \citenamefont {Arvind}}]{shruti-pra-15}%
  \BibitemOpen
  \bibfield  {author} {\bibinfo {author} {\bibfnamefont {S.}~\bibnamefont
  {Dogra}}, \bibinfo {author} {\bibfnamefont {K.}~\bibnamefont {Dorai}}, \ and\
  \bibinfo {author} {\bibnamefont {Arvind}},\ }\href@noop {} {\bibfield
  {journal} {\bibinfo  {journal} {Phys. Rev. A},\ }\textbf {\bibinfo {volume}
  {91}},\ \bibinfo {pages} {022312} (\bibinfo {year} {2015})}\BibitemShut
  {NoStop}%
\bibitem [{\citenamefont {Das}\ \emph {et~al.}(2015)\citenamefont {Das},
  \citenamefont {Dogra}, \citenamefont {Dorai},\ and\ \citenamefont
  {Arvind}}]{debmalya-pra-15}%
  \BibitemOpen
  \bibfield  {author} {\bibinfo {author} {\bibfnamefont {D.}~\bibnamefont
  {Das}}, \bibinfo {author} {\bibfnamefont {S.}~\bibnamefont {Dogra}}, \bibinfo
  {author} {\bibfnamefont {K.}~\bibnamefont {Dorai}}, \ and\ \bibinfo {author}
  {\bibnamefont {Arvind}},\ }\href@noop {} {\bibfield  {journal} {\bibinfo
  {journal} {Phys. Rev. A},\ }\textbf {\bibinfo {volume} {92}},\ \bibinfo
  {pages} {022307} (\bibinfo {year} {2015})}\BibitemShut {NoStop}%
\bibitem [{\citenamefont {Kawamura}\ \emph {et~al.}(2006)\citenamefont
  {Kawamura}, \citenamefont {Morimoto}, \citenamefont {Mori}, \citenamefont
  {Sawae}, \citenamefont {Takarabe},\ and\ \citenamefont
  {Manmoto}}]{kawamura-ijqc-06}%
  \BibitemOpen
  \bibfield  {author} {\bibinfo {author} {\bibfnamefont {M.}~\bibnamefont
  {Kawamura}}, \bibinfo {author} {\bibfnamefont {T.}~\bibnamefont {Morimoto}},
  \bibinfo {author} {\bibfnamefont {Y.}~\bibnamefont {Mori}}, \bibinfo {author}
  {\bibfnamefont {R.}~\bibnamefont {Sawae}}, \bibinfo {author} {\bibfnamefont
  {K.}~\bibnamefont {Takarabe}}, \ and\ \bibinfo {author} {\bibfnamefont
  {Y.}~\bibnamefont {Manmoto}},\ }\href@noop {} {\bibfield  {journal} {\bibinfo
   {journal} {Int. J. Qtm. Chem.},\ }\textbf {\bibinfo {volume} {106}},\
  \bibinfo {pages} {3108} (\bibinfo {year} {2006})}\BibitemShut {NoStop}%
\bibitem [{\citenamefont {Singh}\ \emph {et~al.}(2014)\citenamefont {Singh},
  \citenamefont {Arvind},\ and\ \citenamefont {Dorai}}]{harpreet-pra-14}%
  \BibitemOpen
  \bibfield  {author} {\bibinfo {author} {\bibfnamefont {H.}~\bibnamefont
  {Singh}}, \bibinfo {author} {\bibnamefont {Arvind}}, \ and\ \bibinfo {author}
  {\bibfnamefont {K.}~\bibnamefont {Dorai}},\ }\href@noop {} {\bibfield
  {journal} {\bibinfo  {journal} {Phys. Rev. A},\ }\textbf {\bibinfo {volume}
  {90}},\ \bibinfo {pages} {052329} (\bibinfo {year} {2014})}\BibitemShut
  {NoStop}%
\bibitem [{\citenamefont {Guhne}\ \emph {et~al.}(2002)\citenamefont {Guhne},
  \citenamefont {Hyllus}, \citenamefont {Bruss}, \citenamefont {Ekert},
  \citenamefont {Lewenstein}, \citenamefont {Macchiavello},\ and\ \citenamefont
  {Sanpera}}]{guhne-pra-02}%
  \BibitemOpen
  \bibfield  {author} {\bibinfo {author} {\bibfnamefont {O.}~\bibnamefont
  {Guhne}}, \bibinfo {author} {\bibfnamefont {P.}~\bibnamefont {Hyllus}},
  \bibinfo {author} {\bibfnamefont {D.}~\bibnamefont {Bruss}}, \bibinfo
  {author} {\bibfnamefont {A.}~\bibnamefont {Ekert}}, \bibinfo {author}
  {\bibfnamefont {M.}~\bibnamefont {Lewenstein}}, \bibinfo {author}
  {\bibfnamefont {C.}~\bibnamefont {Macchiavello}}, \ and\ \bibinfo {author}
  {\bibfnamefont {A.}~\bibnamefont {Sanpera}},\ }\href@noop {} {\bibfield
  {journal} {\bibinfo  {journal} {Phys. Rev. A},\ }\textbf {\bibinfo {volume}
  {66}},\ \bibinfo {pages} {062305} (\bibinfo {year} {2002})}\BibitemShut
  {NoStop}%
\bibitem [{\citenamefont {Guhne}\ and\ \citenamefont
  {Hyllus}(2003)}]{guhne-ijtp-03}%
  \BibitemOpen
  \bibfield  {author} {\bibinfo {author} {\bibfnamefont {O.}~\bibnamefont
  {Guhne}}\ and\ \bibinfo {author} {\bibfnamefont {P.}~\bibnamefont {Hyllus}},\
  }\href@noop {} {\bibfield  {journal} {\bibinfo  {journal} {Int. J. Theor.
  Phys.},\ }\textbf {\bibinfo {volume} {42}},\ \bibinfo {pages} {1001}
  (\bibinfo {year} {2003})}\BibitemShut {NoStop}%
\bibitem [{\citenamefont {Toth}\ and\ \citenamefont
  {Guhne}(2005)}]{toth-prl-05}%
  \BibitemOpen
  \bibfield  {author} {\bibinfo {author} {\bibfnamefont {G.}~\bibnamefont
  {Toth}}\ and\ \bibinfo {author} {\bibfnamefont {O.}~\bibnamefont {Guhne}},\
  }\href@noop {} {\bibfield  {journal} {\bibinfo  {journal} {Phys. Rev.
  Lett.},\ }\textbf {\bibinfo {volume} {94}},\ \bibinfo {pages} {060501}
  (\bibinfo {year} {2005})}\BibitemShut {NoStop}%
\bibitem [{\citenamefont {Guhne}\ and\ \citenamefont
  {Seevinck}(2010)}]{guhne-njp-10}%
  \BibitemOpen
  \bibfield  {author} {\bibinfo {author} {\bibfnamefont {O.}~\bibnamefont
  {Guhne}}\ and\ \bibinfo {author} {\bibfnamefont {M.}~\bibnamefont
  {Seevinck}},\ }\href@noop {} {\bibfield  {journal} {\bibinfo  {journal} {New
  J. Phys.},\ }\textbf {\bibinfo {volume} {12}},\ \bibinfo {pages} {053002}
  (\bibinfo {year} {2010})}\BibitemShut {NoStop}%
\bibitem [{\citenamefont {Szangolies}\ \emph {et~al.}(2015)\citenamefont
  {Szangolies}, \citenamefont {Kampermann},\ and\ \citenamefont
  {Bruss}}]{dagmar-njp-15}%
  \BibitemOpen
  \bibfield  {author} {\bibinfo {author} {\bibfnamefont {J.}~\bibnamefont
  {Szangolies}}, \bibinfo {author} {\bibfnamefont {H.}~\bibnamefont
  {Kampermann}}, \ and\ \bibinfo {author} {\bibfnamefont {D.}~\bibnamefont
  {Bruss}},\ }\href@noop {} {\bibfield  {journal} {\bibinfo  {journal} {New J.
  Phys.},\ }\textbf {\bibinfo {volume} {17}},\ \bibinfo {pages} {113051}
  (\bibinfo {year} {2015})}\BibitemShut {NoStop}%
\bibitem [{\citenamefont {MATLAB}(2015)}]{matlab-book}%
  \BibitemOpen
  \bibfield  {author} {\bibinfo {author} {\bibnamefont {MATLAB}},\ }\href@noop
  {} {\emph {\bibinfo {title} {Version 8.5.0 (R2015a)}}}\ (\bibinfo
  {publisher} {MathWorks Inc.},\ \bibinfo {address} {Natick, Massachusetts},\
  \bibinfo {year} {2015})\BibitemShut {NoStop}%
\bibitem [{\citenamefont {Sturm}(1999)}]{sturm-oms-99}%
  \BibitemOpen
  \bibfield  {author} {\bibinfo {author} {\bibfnamefont {J.~F.}\ \bibnamefont
  {Sturm}},\ }\href@noop {} {\bibfield  {journal} {\bibinfo  {journal} {Opt.
  Meth. Soft.},\ }\textbf {\bibinfo {volume} {11}},\ \bibinfo {pages} {625}
  (\bibinfo {year} {1999})}\BibitemShut {NoStop}%
\bibitem [{\citenamefont {Ernst}\ \emph {et~al.}(1990)\citenamefont {Ernst},
  \citenamefont {Bodenhausen},\ and\ \citenamefont {Wokaun}}]{ernst-book}%
  \BibitemOpen
  \bibfield  {author} {\bibinfo {author} {\bibfnamefont {R.~R.}\ \bibnamefont
  {Ernst}}, \bibinfo {author} {\bibfnamefont {G.}~\bibnamefont {Bodenhausen}},
  \ and\ \bibinfo {author} {\bibfnamefont {A.}~\bibnamefont {Wokaun}},\
  }\href@noop {} {\emph {\bibinfo {title} {Principles of Nuclear Magnetic
  Resonance in One and Two Dimensions (International Series of Monographs on
  Chemistry)}}}\ (\bibinfo  {publisher} {Clarendon Press},\ \bibinfo {address}
  {Oxford, United Kingdom},\ \bibinfo {year} {1990})\BibitemShut {NoStop}%
\bibitem [{\citenamefont {Cory}\ \emph {et~al.}(1998)\citenamefont {Cory},
  \citenamefont {Price},\ and\ \citenamefont {Havel}}]{cory-physicad}%
  \BibitemOpen
  \bibfield  {author} {\bibinfo {author} {\bibfnamefont {D.}~\bibnamefont
  {Cory}}, \bibinfo {author} {\bibfnamefont {M.}~\bibnamefont {Price}}, \ and\
  \bibinfo {author} {\bibfnamefont {T.}~\bibnamefont {Havel}},\ }\href@noop {}
  {\bibfield  {journal} {\bibinfo  {journal} {Physica D},\ }\textbf {\bibinfo
  {volume} {120}},\ \bibinfo {pages} {82} (\bibinfo {year} {1998})}\BibitemShut
  {NoStop}%
\bibitem [{\citenamefont {Uhlmann}(1976)}]{uhlmann-rpmp-76}%
  \BibitemOpen
  \bibfield  {author} {\bibinfo {author} {\bibfnamefont {A.}~\bibnamefont
  {Uhlmann}},\ }\href@noop {} {\bibfield  {journal} {\bibinfo  {journal} {Rep.
  Math. Phys.},\ }\textbf {\bibinfo {volume} {9}},\ \bibinfo {pages} {273}
  (\bibinfo {year} {1976})}\BibitemShut {NoStop}%
\bibitem [{\citenamefont {Jozsa}(1994)}]{jozsa-jmo-94}%
  \BibitemOpen
  \bibfield  {author} {\bibinfo {author} {\bibfnamefont {R.}~\bibnamefont
  {Jozsa}},\ }\href@noop {} {\bibfield  {journal} {\bibinfo  {journal} {J. Mod.
  Opt.},\ }\textbf {\bibinfo {volume} {41}},\ \bibinfo {pages} {2315} (\bibinfo
  {year} {1994})}\BibitemShut {NoStop}%
\bibitem [{\citenamefont {Leskowitz}\ and\ \citenamefont
  {Mueller}(2004)}]{leskowitz-pra-04}%
  \BibitemOpen
  \bibfield  {author} {\bibinfo {author} {\bibfnamefont {G.~M.}\ \bibnamefont
  {Leskowitz}}\ and\ \bibinfo {author} {\bibfnamefont {L.~J.}\ \bibnamefont
  {Mueller}},\ }\href@noop {} {\bibfield  {journal} {\bibinfo  {journal} {Phys.
  Rev. A},\ }\textbf {\bibinfo {volume} {69}},\ \bibinfo {pages} {052302}
  (\bibinfo {year} {2004})}\BibitemShut {NoStop}%
\bibitem [{\citenamefont {Chi}\ and\ \citenamefont {Lee}(2003)}]{chi-jpamg-03}%
  \BibitemOpen
  \bibfield  {author} {\bibinfo {author} {\bibfnamefont {D.~P.}\ \bibnamefont
  {Chi}}\ and\ \bibinfo {author} {\bibfnamefont {S.}~\bibnamefont {Lee}},\
  }\href@noop {} {\bibfield  {journal} {\bibinfo  {journal} {J. Phys. A: Math.
  Gen.},\ }\textbf {\bibinfo {volume} {36}},\ \bibinfo {pages} {11503}
  (\bibinfo {year} {2003})}\BibitemShut {NoStop}%
\bibitem [{\citenamefont {Jami}\ and\ \citenamefont
  {Sarbishei}(2005)}]{jami-ind-05}%
  \BibitemOpen
  \bibfield  {author} {\bibinfo {author} {\bibfnamefont {S.}~\bibnamefont
  {Jami}}\ and\ \bibinfo {author} {\bibfnamefont {M.}~\bibnamefont
  {Sarbishei}},\ }\href@noop {} {\bibfield  {journal} {\bibinfo  {journal}
  {Indian J. Phys.},\ }\textbf {\bibinfo {volume} {79}},\ \bibinfo {pages}
  {167} (\bibinfo {year} {2005})}\BibitemShut {NoStop}%
\bibitem [{\citenamefont {Gell-Mann}\ and\ \citenamefont
  {Neeman}(1964)}]{gellmann-book}%
  \BibitemOpen
  \bibfield  {author} {\bibinfo {author} {\bibfnamefont {M.}~\bibnamefont
  {Gell-Mann}}\ and\ \bibinfo {author} {\bibfnamefont {Y.}~\bibnamefont
  {Neeman}},\ }\href@noop {} {\emph {\bibinfo {title} {The Eightfold Way}}}\
  (\bibinfo  {publisher} {W. A. Benjamin Inc.},\ \bibinfo {address} {New
  York},\ \bibinfo {year} {1964})\BibitemShut {NoStop}%
\bibitem [{\citenamefont {Bourennane}\ \emph {et~al.}(2004)\citenamefont
  {Bourennane}, \citenamefont {Eibl}, \citenamefont {Kurtsiefer}, \citenamefont
  {Gaertner}, \citenamefont {Weinfurter}, \citenamefont {Guhne}, \citenamefont
  {Hyllus}, \citenamefont {Bruss}, \citenamefont {Lewenstein},\ and\
  \citenamefont {Sanpera}}]{bourennane-prl-04}%
  \BibitemOpen
  \bibfield  {author} {\bibinfo {author} {\bibfnamefont {M.}~\bibnamefont
  {Bourennane}}, \bibinfo {author} {\bibfnamefont {M.}~\bibnamefont {Eibl}},
  \bibinfo {author} {\bibfnamefont {C.}~\bibnamefont {Kurtsiefer}}, \bibinfo
  {author} {\bibfnamefont {S.}~\bibnamefont {Gaertner}}, \bibinfo {author}
  {\bibfnamefont {H.}~\bibnamefont {Weinfurter}}, \bibinfo {author}
  {\bibfnamefont {O.}~\bibnamefont {Guhne}}, \bibinfo {author} {\bibfnamefont
  {P.}~\bibnamefont {Hyllus}}, \bibinfo {author} {\bibfnamefont
  {D.}~\bibnamefont {Bruss}}, \bibinfo {author} {\bibfnamefont
  {M.}~\bibnamefont {Lewenstein}}, \ and\ \bibinfo {author} {\bibfnamefont
  {A.}~\bibnamefont {Sanpera}},\ }\href@noop {} {\bibfield  {journal} {\bibinfo
   {journal} {Phys. Rev. Lett.},\ }\textbf {\bibinfo {volume} {92}},\ \bibinfo
  {pages} {087902} (\bibinfo {year} {2004})}\BibitemShut {NoStop}%
\bibitem [{\citenamefont {Brandao}(2005)}]{brandao-pra-05}%
  \BibitemOpen
  \bibfield  {author} {\bibinfo {author} {\bibfnamefont {F.~G. S.~L.}\
  \bibnamefont {Brandao}},\ }\href@noop {} {\bibfield  {journal} {\bibinfo
  {journal} {Phys. Rev. A},\ }\textbf {\bibinfo {volume} {72}},\ \bibinfo
  {pages} {022310} (\bibinfo {year} {2005})}\BibitemShut {NoStop}%
\end{thebibliography}
\end{document}